\documentclass{aastex62}
\usepackage{amsmath}
\usepackage{units}

\usepackage{amsmath}
\usepackage{csquotes}

\usepackage{xspace}
\newcommand{\kms}     {\,km\,s$^{-1}$\xspace}
\newcommand{\mjy}     {\,mJy\,beam$^{-1}$\xspace}

\def \sun {$_{\odot}$}

\graphicspath{{./}{figures/}}

\received{\today}
\revised{ }
\accepted{ }
\submitjournal{ApJ}

\shorttitle{Grain alignment in PDRs}
\shortauthors{Soam et al.}
\begin{document}

%\title{Do collisions control the dust grain disalignment?}
\title{On the collisional disalignment of dust grains in illuminated and shaded regions of IC\,63}

\correspondingauthor{Archana Soam, B-G Andersson}
\email{asoam@usra.edu, bgandersson@sofia.usra.edu}

\author[0000-0002-6386-2906]{Archana Soam}
\affiliation{SOFIA Science Center, USRA, NASA Ames Research Center, M.S.-12, N232, Moffett Field, CA 94035, USA}

\author[0000-0001-6717-0686]{B-G Andersson}
\affil{SOFIA Science Center, USRA, NASA Ames Research Center, M.S.-12, N232, Moffett Field, CA 94035, USA}

\author{Jose Acosta-Pulido}
\affiliation{Instituto de Astrof\'isica de Canarias (IAC), C/O Via Lactea, s/n E38205- La Laguna, Tenerife, Spain}

\author{Manuel Fern\'andez L\'opez}
\affiliation{Instituto Argentino de Radioastronom\'ia (CCT La Plata, CONICET), C.C.5, (1984) Villa Elisa, Buenos Aires, Argentina}

\author{John E. Vaillancourt}
\affiliation{Lincoln Laboratory, Massachusetts Institute of Technology, 244 Wood St., Lexington, MA 02421, USA}

\author{Susanna L. Widicus Weaver}
\affiliation{Department of Chemistry, Emory University, Atlanta, GA 30322}

\author{Vilppu Piirola}
\affiliation{Department of Physics and Astronomy, University of Turku, FI-20014 Turku, Finland}
\affiliation{Finnish Centre for Astronomy with ESO, University of Turku, Finland}

\author{Michael S. Gordon}
\affiliation{SOFIA Science Center, USRA, NASA Ames Research Center, M.S.-12, N232, Moffett Field, CA 94035, USA}



\begin{abstract}

Interstellar dust grain alignment causes polarization from UV to mm wavelengths, allowing the study of the geometry and strength of the magnetic field. Over last couple of decades observations and theory have led to the establishment of the Radiative Alignment Torque (RAT) mechanism as leading candidate to explain the effect. With a quantitatively well constrained theory, polarization can be used not only to study the interstellar magnetic field, but also the dust and other environmental parameters. Photo-dissociation Regions (PDRs), with their intense, anisotropic radiation fields, consequent rapid $\rm H_{2}$ formation, and high spatial density-contrast provide a rich environment for such studies. Here we discuss an expanded optical, NIR, and mm-wave study of the IC\,63 nebula, showing strong $\rm H_{2}$ formation-enhanced alignment and the first direct empirical evidence for disalignment due to gas-grain collisions using high-resolution $\rm HCO^{+}$(J=1-0) observations. We find that relative amount of polarization is marginally anti-correlated with column density of $\rm HCO^{+}$.  However, separating the lines of sight of optical polarimetry into those behind, or in front of, a dense clump as seen from $\gamma$ Cas, the distribution separates into two well defined sets, with data corresponding to \enquote{shaded} gas having a shallower slope.  This is expected if the decrease in polarization is caused by collisions since collisional disalignment rate is proportional to R$_C\propto n\sqrt{T}$.  Ratios of the best-fit slopes for the \enquote{illuminated} and \enquote{shaded} samples of lines of sight agrees, within the uncertainties, with the square-root of the two-temperature H$_2$ excitation in the nebula seen by \citet{2009MNRAS.400..622T}.

\end{abstract}

\keywords{interstellar dust, polarimetry}


\section{Introduction} \label{sec:intro}
Magnetic fields play a crucial role in the evolution of the interstellar medium (ISM; \citealt{2004Ap&SS.292..225C}). However, quantitative determinations of the field are notoriously difficult and often very resource intensive to make, such as for Zeeman effect observations \citep{2008ApJ...680..457T}. The easiest observational method for characterizing the magnetic field is via broad-band polarimetry in either the Ultraviolet/Optical/Near Infrared (UV/O/NIR) or the Far Infrared/[sub]mm-wave (FIR/mm). Such polarization is due to dichroic extinction (UV/O/NIR) or emission (FIR) by asymmetric dust grains aligned usually with the magnetic field.  To reliably derive magnetic field characteristics from polarization observations, e.g. via \enquote{Davis-Chandrasekhar-Fermi analysis} \citep{Davis1951,1953ApJ...118..113C}, we must understand the physics of grain alignment in detail. To do so requires a quantitative understanding of the mechanisms both driving and damping the alignment \citep{1998ApJ...508..157D}. 

Significant progress has recently been made in the understanding of interstellar grain alignment, both theoretically and observationally. Theoretically, after realizing that paramagnetic relaxation \citep{1951ApJ...114..206D, 1979ApJ...231..404P, 1986ApJ...308..281M} cannot align grains due to internal grain excitation \citep{1999ApJ...520L..67L}, a new theory was developed based on radiative drivers \citep{1976Ap&SS..43..291D,1996ApJ...470..551D,2007MNRAS.378..910L}. Observationally, paramagnetic relaxation alignment was challenged \citep{jones1967} by the finding of grain alignment at large opacities \citep{jones1984, hough2008}. In parallel, the \enquote{Radiative Alignment Torque} (RAT) theory has now faced, and met, a number of tests. \citet{2007ApJ...665..369A} and \citet{2008ApJ...674..304W}, and recently \citet{medan2019}, showed that the alignment is directly correlated with the radiation field strength. \citet{2010ApJ...720.1045A} showed that the alignment around a localized source (HD 97300, in Chameleon) falls off with distance to the star and that \citep[cf][]{2011A&A...534A..19A} this alignment varies with the angle between the radiation and the magnetic fields consistent with specific RAT theory predictions. \citet{2013ApJ...775...84A} have shown that the grain alignment in the reflection nebula IC\ 63 depends on the $\rm H_{2}$ formation rate, again consistent with RAT theory predictions as demonstrated in \citet{2015MNRAS.448.1178H}. For a recent review of RAT alignment, see \citet{2015ARA&A..53..501A}.

Balancing the alignment, several mechanisms might contribute to grain randomization \citep{1998ApJ...508..157D}. For the neutral medium, gas-grain collisions are thought to dominate. In this case, it is easy to show that the characteristic time for the disalignment is given by $\rm t_{random} \propto (\textit{a}\cdot n_{gas} \sqrt{T_{gas}})^{-1}$ \citep{whittet2003}, where $\rm n_{gas}$ and $\rm T_{gas}$ are the gas density and temperature and \textit{a} is the effective grain radius. 

We can understand the influence of the driving and damping on the grain alignment through RAT theory.  Here we will only sketch a mostly phenomenological approach.  We refer the interested reader to \citet{2015MNRAS.448.1178H} for a more quantitative analysis.  Following \citet{2015MNRAS.448.1178H} we can write the angular momentum (\textbf{J}) evolution of a dust grain of selected size as the balance between \enquote{driving} torques and \enquote{damping} by gas-grain collisions and far infrared emission by the grains:

\begin{equation}
d\textbf{J}/dt = {\bf \Gamma} - C_{coll} (1+F_{IR}) n_H \sqrt{T_{gas}}\,\,\textbf{J}
\end{equation}

\noindent
where ${\bf \Gamma}$ is the external regular torques on the grain and C$_{coll}$ is a constant for a given grain.  The expression  (1+F$_{IR}$) contains the correction to the angular momentum damping from infrared emission, which, however, is expected to be small for grains larger than $\sim$0.1$\mu$m \citep{1998ApJ...508..157D}.

The external torque $\Gamma$ is made up of a direct RAT component and a H$_2$ induced torque.  These are not fully independent effects in the complex, non-linear alignment dynamics, where the H$_2$ torques enhance the radiatively driven torques, lifting grains that would otherwise achieve only poor alignment to efficient alignment at \enquote{high-J attractor points} in the dynamical phase-space diagrams \citep{2007MNRAS.378..910L,2015MNRAS.448.1178H}.   For simplicity, we will, however here, treat them as separate terms.  The radiative torques for grains larger than the minimal available wavelength are proportional to

\begin{equation}
\Gamma_{Rad} = C_{RAT} <\lambda> u_{rad}
\end{equation}

\noindent
where u$_{rad}$ is the radiation field strength with an average wavelength of $<\lambda>$ and C$_{RAT}$ contains both general constants and the geometry - including radiation field anisotropy - of the environment.  The H$_2$ formation torques are given by

\begin{equation}
\Gamma_{H_2} = C_{H_2} \gamma_H n_H (1-y) \sqrt{T_{gas}}
\end{equation}

\noindent
where $\gamma_H$ is the conversion efficiency from atomic
to molecular hydrogen, y = 2n(H$_2$)/n$_H$ is the fraction of molecular hydrogen, which is controlled by the radiation field at $\lambda<1107$\AA \,\,and the self-shielding characteristics of the hydrogen molecule \citep{federman1979, 1986ApJS...62..109V}, and we have, similarly to the radiative torques incorporated the H$_2$ formation parameters (the grain shape, the number of chemisorption sites on the grain, etc.) in the factor C$_{H_2}$.

In steady state d\textbf{J}/dt = 0 and therefore the maximum angular momentum of the grain in steady state will be \citep[cf][]{2015MNRAS.448.1178H}:

\begin{equation}
J_{max} = \frac{ \nicefrac{C_{H_2}}{C_{coll}}} {1+F_{IR}} \gamma_H (1-y) + \frac{ \nicefrac{C_{RAT}}{C_{coll}}} {(1+F_{IR})(n_H \sqrt{T_{gas}})}
\label{balanceequ}
\end{equation}

\noindent
which, ultimately, controls the polarization efficiency, albeit dependent on a number of complex processes and with several poorly known parameters, such as the grain axis ratio, the amount of super-paramagnetism of the grains, etc.   As shown by \citet{2015MNRAS.448.1178H}, taking the full range of physics into account, \textit{ab inito} modeling based on RAT theory can  reproduce the over-all characteristics of the observed polarization in IC\,63.  We will not, here, try to make such detailed modeling of the alignment, but note that the equilibrium value of J$_{max}$ depends inversely on the gas-grain collision rate.

However, gas-grain collisional damping has, to date, not been directly observationally confirmed (c.f. \citealt{2007ApJ...665..369A}).  Here we expand on our earlier results and probe the role of the collisional disalignment.  In the present analysis we employ a perturbation-style analysis and will analyze the deviation from the average alignment.  That is, as discussed in section \ref{sec:diss}, we fit the over-all polarization dependence on the H$_2$ formation rate, find the offsets from this average fit and analyze these difference in the polarization ($\Delta$P) values in terms of the gas density to probe for local, differential collisional effects.

IC\,63 is an excellent target for studying radiation driven alignment and collisional disalignment. The region is near-by (at d$\sim$200 pc; \citealt{vanleeuwen2007, 2013ApJ...775...84A}) and well characterized with (1) the radiation field from a B0 IV star, $\gamma$ Cas \citep{2005AJ....129..954K, 2005ApJ...628..750F}, (2) the cloud has been well studied and modeled on global scales \citep{1995A&A...302..223J, 2015MNRAS.448.1178H, 2018A&A...619A.170A}, (3) a number of background targets are imaged with polarimetry \citep{2017MNRAS.465..559S} (4) being of limited angular size and relatively dense, molecular rotation lines are used to study spatial variations in collision rates with moderate investments in telescope time. In addition to the generally high densities and temperatures \citep{1994A&A...282..605J}, the compression ridge region of the nebula shows additional enhancements in both, with density $\rm n = 10^{5} - 10^{6} cm^{-3}$ \citep{2005IAUS..231P.148P}, and temperature T$\sim$ 600\,K \citep{2010ApJ...725..159F}.

The IC\,63 cloud has been studied in O/NIR polarimetry by \citet{2013ApJ...775...84A} and \citet{2017MNRAS.465..559S}.  The former used optical polarimetry combined with NIR observations of the fluorescence in the 1-0 S(1) line of H$_2$ to show that the grain alignment is enhanced by the formation torques of the ejection from newly formed H$_2$ molecules.  \citet{2015MNRAS.448.1178H} used \textit{ab initio} modeling based on RAT theory to reproduce the grain alignment in the different parts of the cloud, taking into account both the direct radiative torques and the torques from H$_2$ formation.  In both studies outliers exist at anomalously low values of the measured polarization.

In addition to finding $\rm H_{2}$ formation driven grain alignment in parts of IC\,63, \citet{2013ApJ...775...84A} also found anomalously inefficient alignment in the compression ridge region as well as a much steeper fall-off in the fractional polarization ($\rm P/A_{V}$) with column density ($\rm P/A_{V} \propto A^{-1.1\pm0.1}$) than in the general ISM ($\rm P/A_{V} \propto A^{-0.52\pm0.07}$); \citealt{2008ApJ...674..304W}). A power-law exponent of -1 indicates that the alignment is generated only in a surface layer of the nebula. Based on the studies by \citet{1994A&A...282..605J, 2005IAUS..231P.148P} and \citet{2010ApJ...725..159F}, the rapid loss of alignment with $\rm A_{V}$ and the high contrast in density between the compression ridge and cloud in the IC\ 63 could likely point to collisional disalignment as a cause of the outliers. 

In this work, we have acquired additional polarimetric measurement of the nebula as well as high-resolution mapping in the $\rm HCO^{+}$ (J=1-0) line, which allow us to address the effects of gas-grain collisions on the grain alignment. The $\rm HCO^{+}$ (J=1-0) line intensity is a sensitive indicator of the gas column density, but only slowly varying functions of temperature \citep{1994A&A...282..605J}. The detailed modeling of Jansen et al. shows that IC\,63 is optically thin ($\rm \tau \sim 0.1$) region due to its relatively warm ($\sim 100\,K$) nature.

The paper is organized as follows: in Section 2, we describe the observations and data reduction; in Sections 3, 4, and 5, we describe, discuss and summarize our results of this work.

\section{Observations} \label{sec:obs}

\subsubsection{Archival Data}

We used our existing optical and NIR polarimetric observations from \citet{2013ApJ...775...84A} and \citet{2017MNRAS.465..559S}. Optical polarization observations towards 12 stars were made at the Nordic Optical Telescope (NOT) on La Palma using ALFOSC and TURPOL instruments in spectro-polarimetry and photo-polarimetry modes, respectively. The details of the observations are given in \citet{2013ApJ...775...84A}.

For correlating the polarization fraction with grain alignment caused by \textit{Purcell mechanism} \citep{1979ApJ...231..404P}, we used the $\rm H_{2}$ 1-0 S(1) observations from \citet{2013ApJ...775...84A}. These deep narrow-band imaging observations at 2.122 $\mu$m were acquired with the WIRCam instrument \citep{2004SPIE.5492..978P} at the Canada$-$France$-$Hawaii Telescope (CFHT). To better analyze the diffuse emission, we had to first perform a star subtraction of the foreground sources. We used the astropy-affliated photutils \citep{Bradley_2019_2533376} package to construct an effective point spread function (ePSF) from stars extracted over the entire image.  Combined with two-dimensional background estimation and modeling, the ePSF was used to subtract away point sources, leaving the underlying diffuse emission intact.

\subsection{Optical spectropolarimetry}

Additional optical spectropolarimetric observations were acquired on 2013, October 18 and 19 from the Intermediate-dispersion Spectrograph and Imaging System (ISIS) mounted at the $\sim$f/11 Cassegrain focus of the 4.2\,m William Herschel Telescope (WHT) in the Canary Islands, Spain. Observations were acquired with both the red and blue sides of the spectrograph. For the blue arm we used the EEV12 CCD providing good quantum efficiency down to the atmospheric cut-off and the R300B grating centered at 4500 $\AA$.  For the red arm we used the Red+ CCD and the R158R grating centered at 6400\,$\AA$. The blue observations were corrupted by scattered light and are not discussed here.  The red data cover $\lambda \approx 0.48 - 0.95\,\mu m$.  For comparison with earlier data we have extracted the polarization values at 0.65 $\mu$m. HD\,204827 was used as the high-polarization standard and HD\,212311 was observed as a zero-polarization standard. We divided the total integration time into  8 half-wave plate (HWP) settings, to minimize the influence of systematic errors. Data reduction was accomplished through standard IRAF routines.

\subsection{$\rm HCO^{+}$ line}
Observations of the J = 1-0 line of HCO$^+$ at 89.189 GHz were conducted using the Combined Array for Research in Millimeter-wave Astronomy (CARMA) in the 15-element mode (six 10.4\,m antennas and nine 6.1~m antennas). Observations were conducted in three tracks separated by a few days (13, 14 and 17 February 2014), with CARMA in its C configuration (baselines ranging 30\,m to 350\,m). A six pointing mosaic with the phase center aiming at $\alpha_{J2000}$=$00^h 59^m 01\fs22$ and $\delta_{J2000}$=$+60\degr 53\arcmin 23\farcs40$ was implemented. Good atmospheric conditions prevailed for the observations with an average opacity at 230\,GHz between 0.15 and 0.3 for the three tracks. The total time on source was 9.4 hours.

The first local oscillator (LO) was set to 91.0163~GHz (3.3~mm). The correlator configuration provided simultaneous observations of the continuum emission (1.9~GHz of total bandwidth) and several lines, including HCO$^+$ (1-0) and HCN (1-0) among others. Here we present the HCO$^+$ (1-0) data (rest frequency 89.18853\,GHz), which lay in one of the ten spectral windows each consisting of 159 channels with 50~KHz each (i.e., 0.16\kms at the observing frequency) and 8~MHz total width (about 26\kms).

The data were edited, calibrated, imaged using the MIRIAD package (\citealt{1995Sault}) in a standard way. Quasars 3C84 and J0102+584 were used as bandpass and phase calibrator, respectively. Observations of Uranus provided the absolute scale for the flux calibration of the dataset. The measured flux of J0102+584 was 2.6\,Jy and the flux uncertainty for CARMA observations at 3\,mm is estimated to be 10\%.

A continuum image was created by combining the dataset of the three tracks using the Miriad program INVERT with a natural robust weighting. The rms noise of the resulting interferomter-only image has a synthesized beam of $5\farcs9\times4\farcs9$ with a P.A. of 48$\degr$ and an rms noise level of 0.3\mjy. No continuum emission is detected over a 3$\sigma$ level. The HCO$^+$ (1-0) line image was created using a natural weighting scheme too and the inteferometer-only velocity cube has a synthesized beam of $6\farcs1\times4\farcs7$ with a P.A. of 57$\degr$ and an rms noise level of 97\mjy.

Single dish observations covering the IC\,63 interferometry pointings were also acquired with the EMIR instrument on the IRAM 30\,m telescope during 2013, July 21-23.  The observations were carried out in on-the-fly (OTF) mapping, position-switching mode, using a velocity resolution of 0.13\kms. The final map covered $4.0\arcmin\times2.8\arcmin$ was obtained.  

After calibrating the CARMA interferometer visibilities, we combined the HCO$^+$ (1-0) with map taken with IRAM\,30\,m. For this purpose we created an interferometer image using a robust weighting of 0.5 and converted the IRAM\,30\,m Dish data to Jy\,beam$^{-1}$ using a Jy-per-K factor of 6.0\footnote{https://www.iram.es/IRAMES/telescope/telescopeSummary} for observations at 86\,GHz. We made a jointly$-$deconvolution of the interferometer and the single-dish datasets using the Miriad program MOSMEM and then restoring the resulting image with a synthesized beam of $5\farcs8\times4\farcs4$ with a P.A. of 50$\degr$. The final image has an rms noise level of 6\,\mjy. Finally, we also made a HCO$^+$ (1-0) integrated intensity image, collapsing the emission from -0.9\,\kms to 1.5\,\kms. This image has an rms noise level of about 24\,\mjy\kms.

\begin{figure}
\gridline{\fig{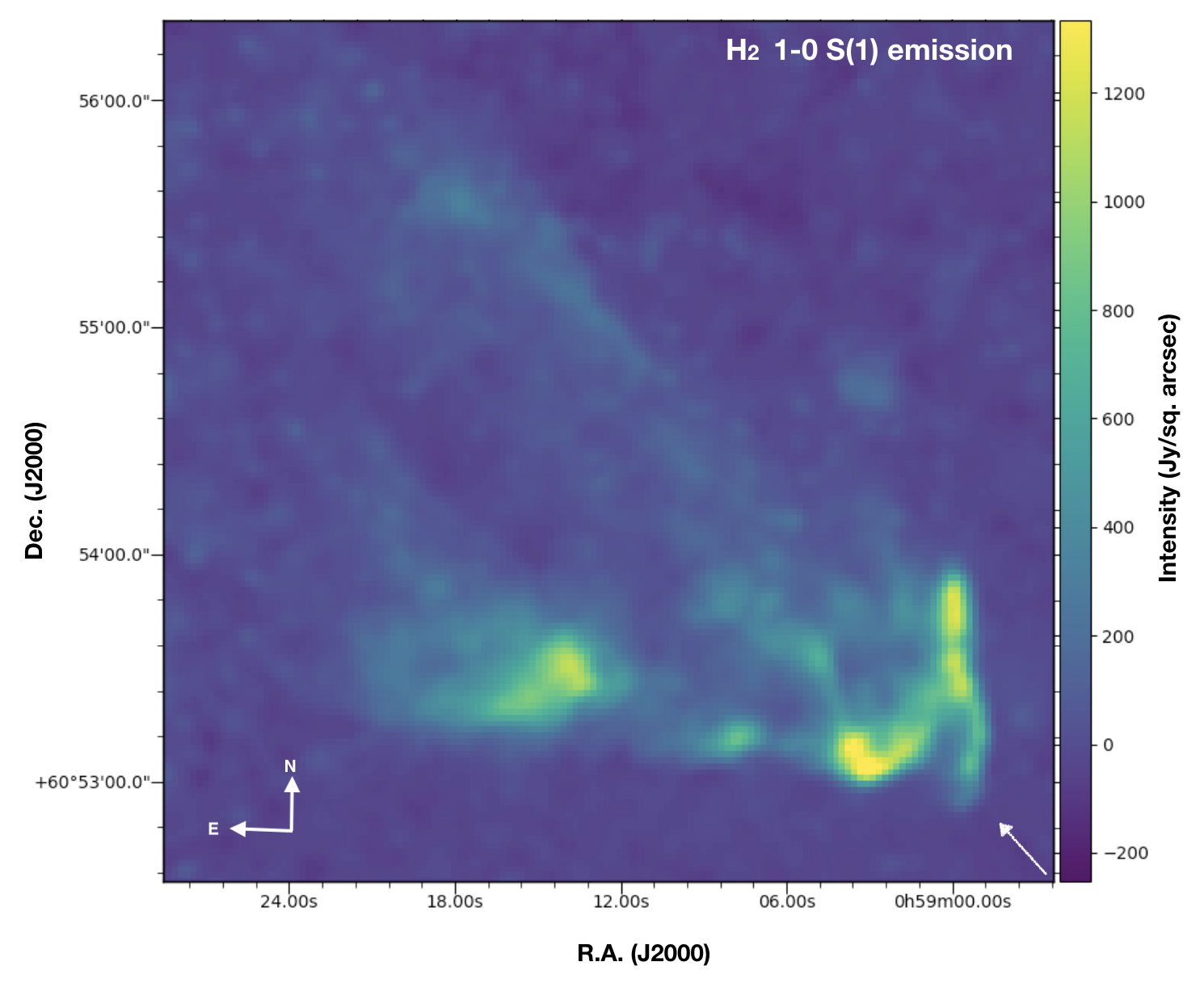}{0.71\textwidth}{(a)}
          }
\gridline{\fig{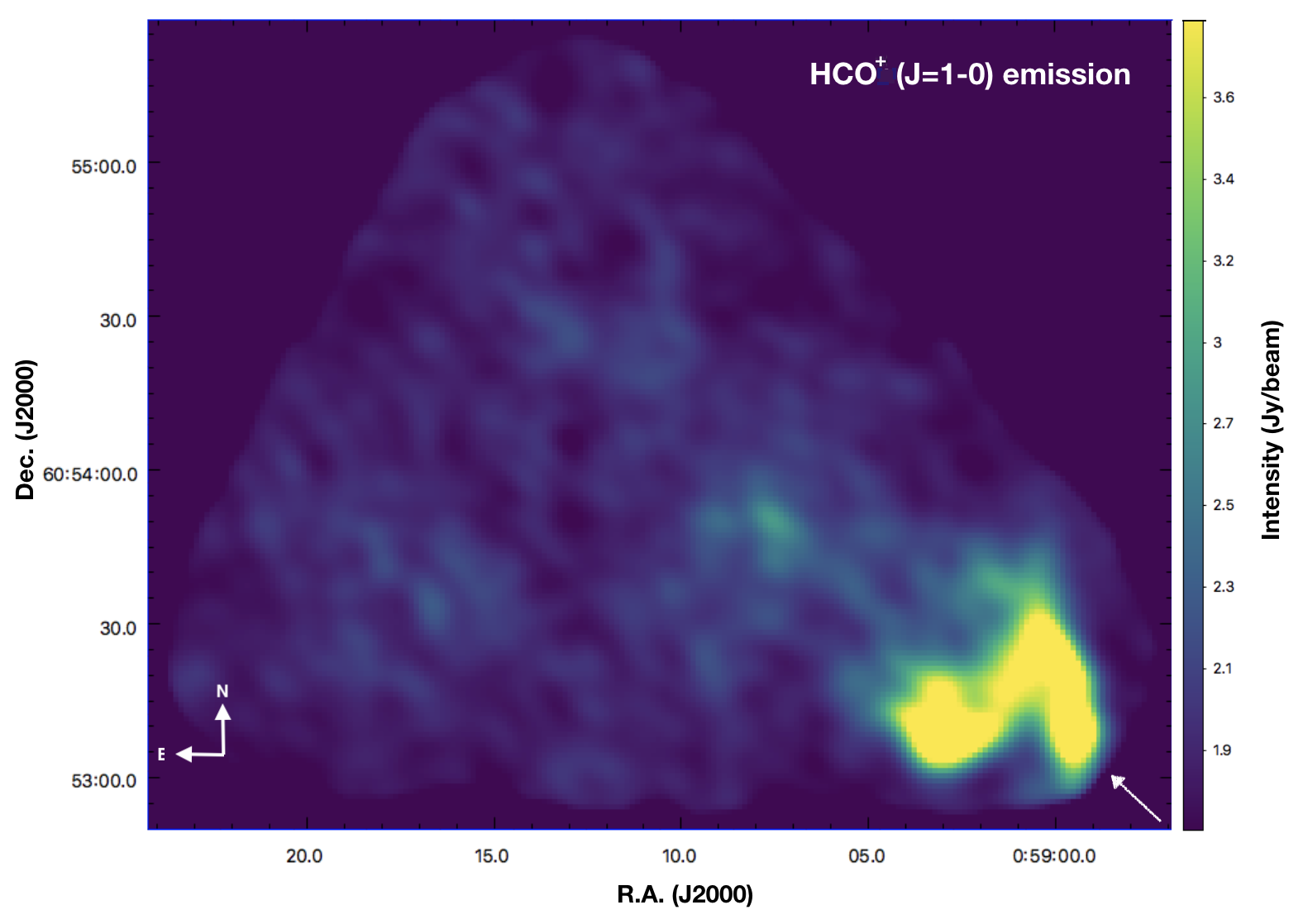}{0.71\textwidth}{(b)}
%          \fig{IC\ 63_HCOp_clumps_stars-eps-converted-to.pdf}{0.3\textwidth}{(e)}
          }
\caption{Upper and lower panels show the $\rm H_{2}$ and $\rm HCO^{+}$ gas emission maps towards IC\ 63 nebula, respectively. The interesting \textit{hook-like} feature can be seen in both images towards the region facing ionizing radiation. The directions of radiation are shown with thick arrow in both the panels. \label{fig:gasemission}}
\end{figure}

\section{Results} \label{sec:res}
Figure \ref{fig:gasemission} shows emission maps of v=1-0 $\rm H_{2}$ J=1-0 (upper panel) and $\rm HCO^{+}$ J=1-0 (lower panel) towards IC\,63. The clumpiness in $\rm H_{2}$ emission with a hook-like emission feature can be clearly seen. Also, the $\rm HCO^{+}$ gas is mostly localized in the front of the nebula looking towards high energy radiation from $\gamma$ Cas (as seen in the PdBI data from \citet{2005IAUS..231P.148P}). The hook-like feature is common in both emission maps indicating the gaseous evaporation on the side facing the ionizing radiation (Caputo et al. 2020, in press). The directions of ionizing radiations (i.e. from south-west to north-east) are shown using white arrows in both panels.

\begin{figure}
\centering
\resizebox{13.0cm}{10.0cm}{\includegraphics{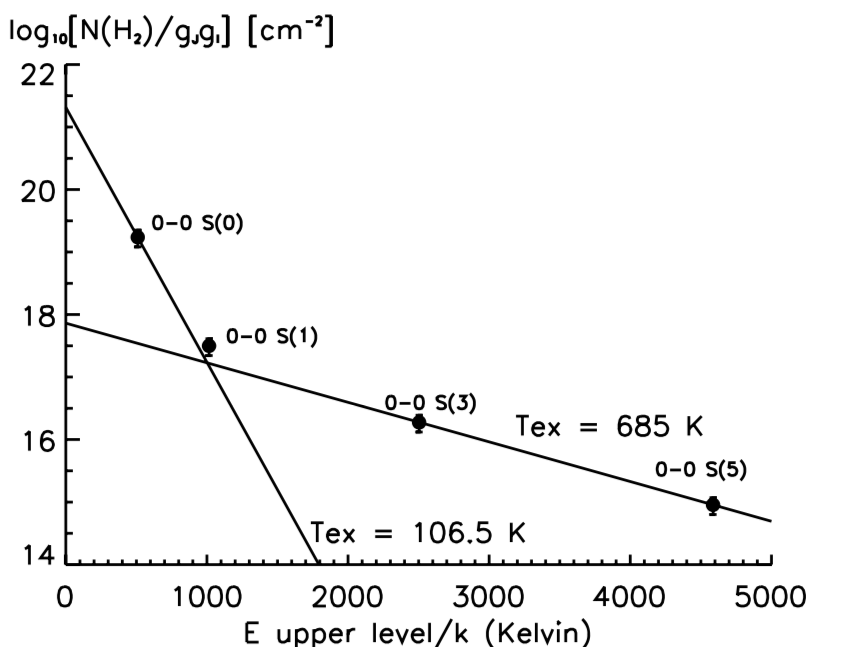}}
\caption{The $\rm H_{2}$ excitation diagram toward the IC\ 63 PDR derived from the ISO-SWS observations. This plot is adopted from \citet{2009MNRAS.400..622T}.}\label{Fig:thi09}
\end{figure}

\citet{2009MNRAS.400..622T} observed the lowest pure-rotational lines S(0); J=2-0, S(1); J=3-1, S(3); J=5-3 and S(5); J=7-5 of $\rm H_{2}$ towards IC\ 63 using the Short Wavelength Spectrometer (SWS) of the \textit{Infrared Space Observatory} (ISO). The apertures of the SWS are fairly large on the scale of the H$_2$ ridge ($20\times27\arcsec$ at S(0), $14\times27\arcsec$ S(1), and $14\times20\arcsec$ at the S(3) and S(5) lines) and the emission may therefore probe several different physical regions. Figure \ref{Fig:thi09} shows their $\rm H_{2}$ excitation diagram toward the IC\,63 PDR and illustrate that the observed intensities are best fit with two thermal components with T=106$\pm$11 and 685$\pm$68\,K, respectively. 
 \citet{2009MNRAS.400..622T} compared their observations with available photo-chemical models \citep{1997ESASP.419..299T, 1999ApJ...527..795K, 2007A&A...467..187R} based on optical absorption and/or millimeter emission data with and without enhanced $\rm H_{2}$ formation rate on grain surfaces. The inferred column densities of these two regions are $\rm (5.9\pm1.8)^{+0.9}_{-0.7} \times 10^{21} cm^{-2}$ and $\rm (1.2\pm0.4)\times 10^{19} cm^{-2}$, corresponding to the warm and hot components, respectively. The ISO temperature estimate of the hot region \citep{2009MNRAS.400..622T} is consistent with the ISOCAM results (630K) of \citet{2005AJ....129..954K} and the \textit{Spitzer/IRS} results of \citet{2010ApJ...725..159F}. Because \citet{2010ApJ...725..159F} did not have access to the long wavelength spectra from $Spitzer$ IRS-LL for IC\,63, they were not able to probe the presence or distribution of the warm ($\sim100$K) gas.

\begin{figure}
\centering
\resizebox{14.5cm}{12.0cm}{\includegraphics{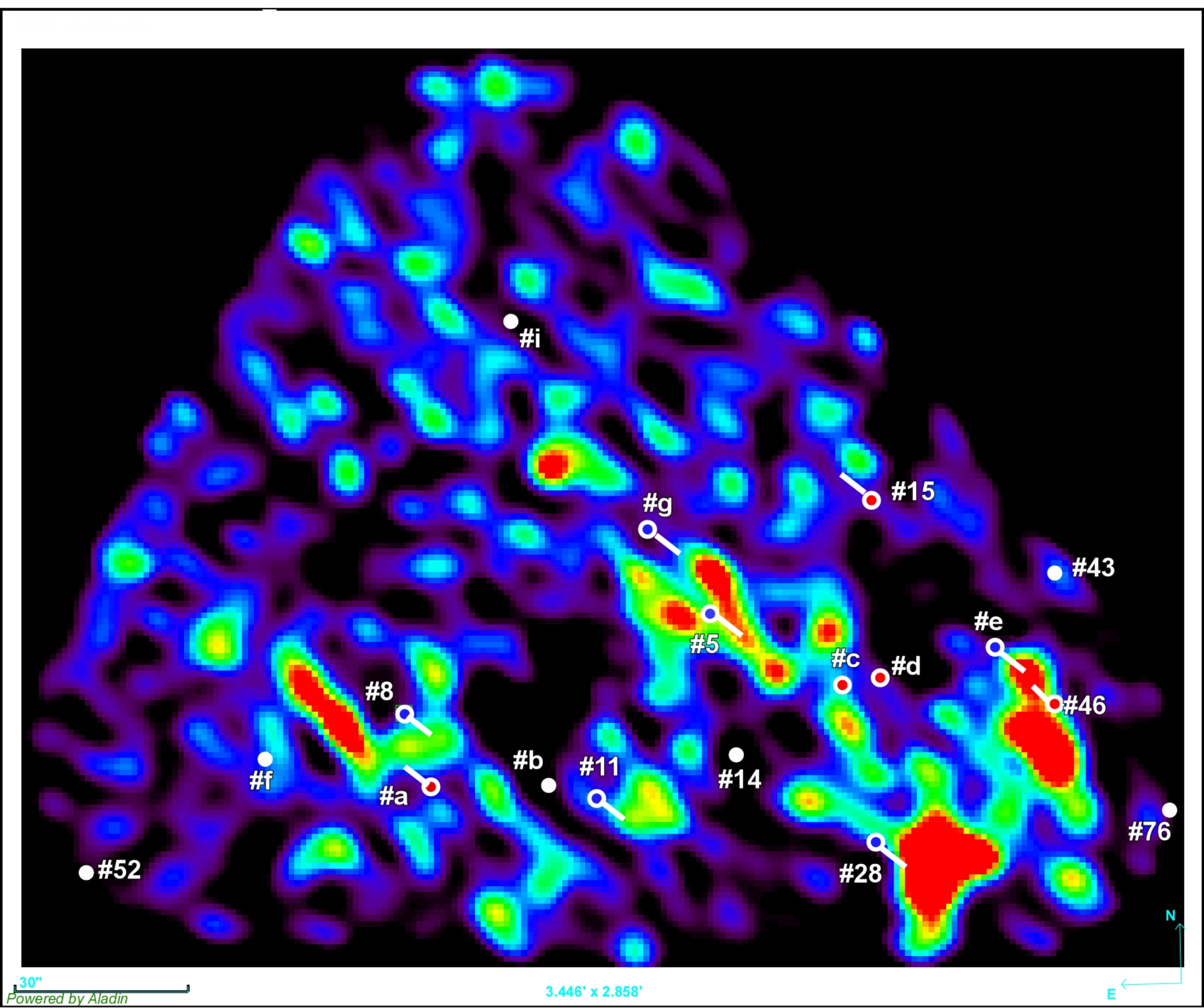}}
\caption{The location of stars with optical polarization observations shown on the $\rm HCO^{+}$(J=1-0) map. Star behind the clumps and in-front of the clumps are shown with blue and red colors, respectively. A number of lines of sight do not clearly fall in either category.  These are plotted as white symbols.}\label{fig:HCO+_clumps}
\end{figure}

\begin{figure}
\centering
\resizebox{16.5cm}{7.0cm}{\includegraphics{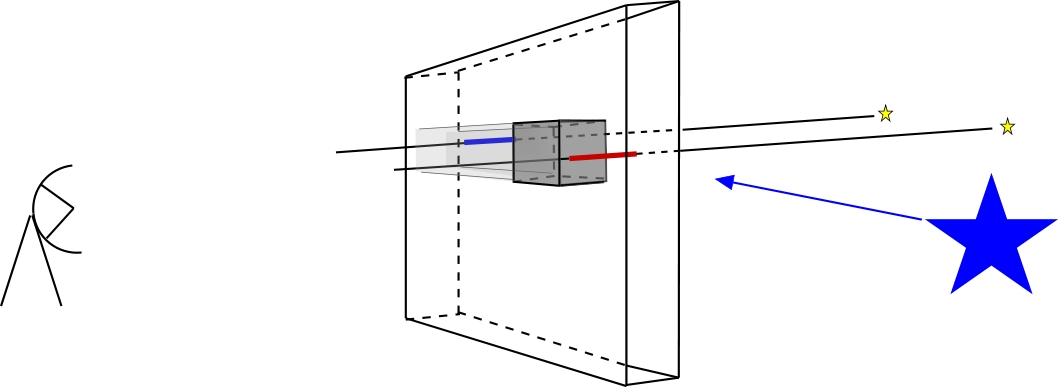}}
\caption{A sketch illustrating the structure of a clumpy cloud with an assumed slab geometry, illuminated by a bright, near-by star ($\gamma$ Cas/ blue star), and probed by the lines of sight to distant (yellow) background stars. A dense clump in the cloud (shaded cuboid) will cast a shadow on the far side of $\gamma$ Cas (light gray region). The gas on a line of sight in front of the clump (red line-segment) will therefore receives a higher radiation dose than gas on a line of sight in the shaded region (blue line-segment).}\label{Fig:cartoon}
\end{figure}

In Figure \ref{fig:HCO+_clumps}, we show the spatial locations of the stars plotted on clumpy $\rm HCO^{+}$ emission towards IC\,63 (this figure is deliberately stretched to see the clumpy structures). The polarization observations of these stars are available in \citet{2013ApJ...775...84A} and \citet{2017MNRAS.465..559S}. The stars are either located, in projection, behind (blue dots) or in front (red dots) of the clumpy structures of $\rm HCO^{+}$ emission. The true, physical, distances of these stars towards IC\,63 are irrelevant as long as we assume that the IC\,63 cloud is the only layer of dust by which the background star light is polarized along the line-of-sight (Soam et al. 2020, submitted). We are only concerned with the locations in the nebulae where the light shining though the cloud and getting polarized, either in front or behind of the $\rm HCO^{+}$ clumps. So the contributions of clumps of the nebulae in polarizing the light is more important than the actual distances of the sources of the light (i.e. background stars). The cartoon shown in Fig. \ref{Fig:cartoon} illustrates this by showing the locations of the main cloud IC\,63 [assuming a slab like geometry], ionizing source with a blue star symbol, and stars behind the cloud shown with yellow star symbols. The $\rm HCO^{+}$ clump is shown as shaded cuboid symbol with its front and behind line-of-sights through the cloud shown as red and blue line-segments. The polarized light from these front and behind the clump line of sights is measured. These $\rm HCO^{+}$ clumps represent the gas densities and the variation of any change in polarization may be due to the dust [in the cloud] and gas collisions.

\begin{figure*}
\gridline{\fig{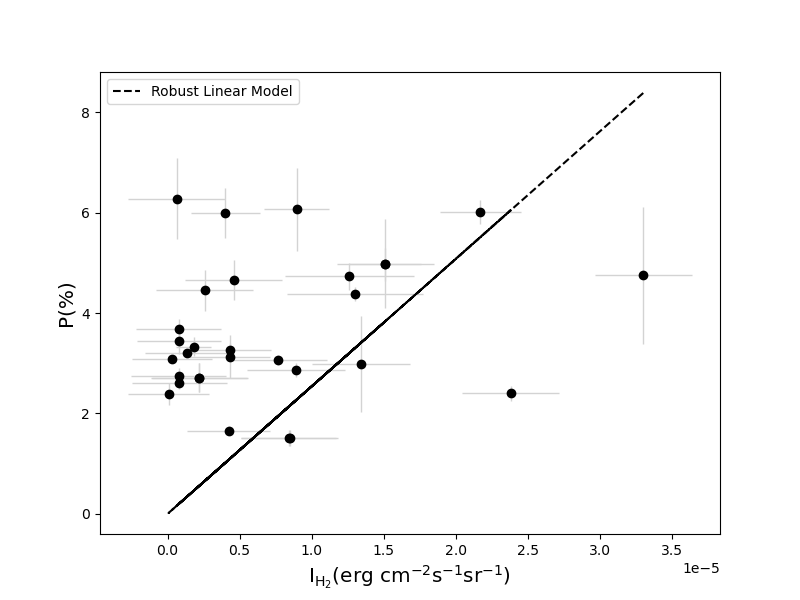}{0.71\textwidth}{(a)}
          }
\gridline{\fig{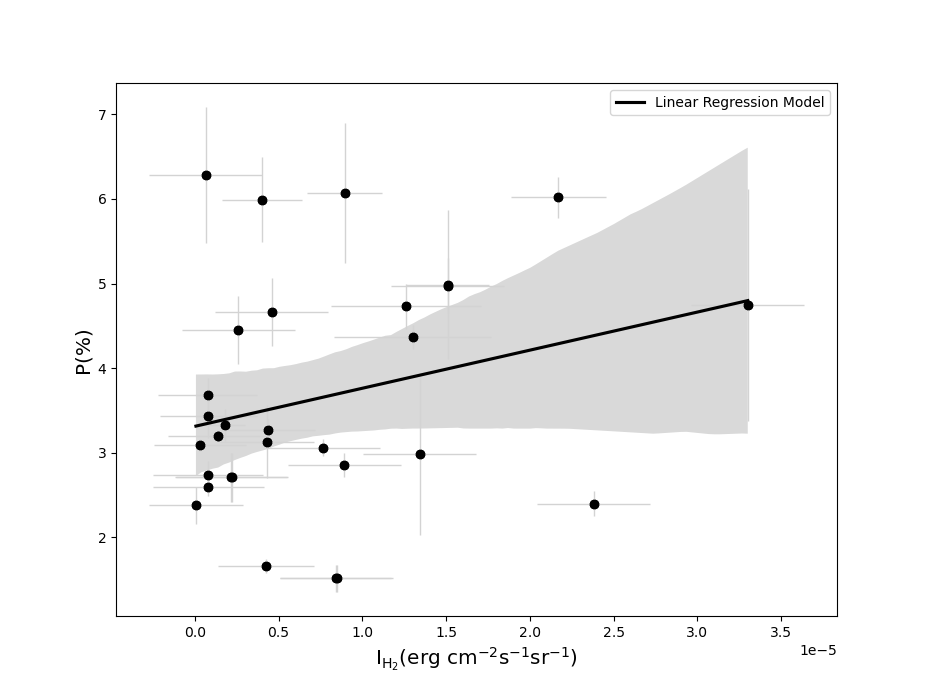}{0.71\textwidth}{(b)}
          }
\caption{{\bf (a)} Distribution of amount of polarization and diffused $\rm H_{2}$ fluorescence with Robust Linear Model fitting. {\bf (b)} Smoother representation of P and $\rm H_{2}$ fluorescence distributions using Kernel Density Estimator (KDE).\label{fig:PvsH2}}
\end{figure*}

Figure \ref{fig:PvsH2} shows the variation of polarization fraction (data includes both from \citet{2013ApJ...775...84A} and \citet{2017MNRAS.465..559S}) with the $\rm H_{2}$ intensities at the locations of nebula where polarization is detected. This variation is fit by using linear regression and robust linear model fits to these data. The upper panel shows the result of the Robust linear model fit. This model is less sensitive to the outliers. The lower panel shows the linear regression fit. Both the models show data fit differently but the output fit value of P is almost the same (i.e. $\sim$3.5\%).

Figure \ref{fig:twotempB} shows the variation of difference in polarization fraction ($\rm \Delta P$) with $\rm HCO^{+}$ intensities at the locations of the stars shown in Figure \ref{fig:HCO+_clumps}\footnote{The number of blue data points in Figure \ref{fig:HCO+_clumps} are six but there are five blue data points shown in Figure \ref{fig:twotempB}. This is because we dropped one star which has unusually high $\rm \Delta P$ and large uncertainty in polarization measurement. There are fourteen red points in Figure \ref{fig:twotempB} but only twelve in Figure \ref{fig:HCO+_clumps}. We considered two extra stars from \citet{2017MNRAS.465..559S} with those from \citet{2013ApJ...775...84A}.}. We denote $\rm \Delta P$ as the difference in observed amount of polarization and best fit value of polarization ($\rm \Delta P = P - P_{fit}$) obtained from Figure \ref{fig:PvsH2}, as $\rm \Delta P$ in this plot. The color of the symbols (red and blue) in Figure \ref{fig:twotempB} are deliberately chosen to connect this figure with locations of stars shown in Figure \ref{fig:HCO+_clumps}. The white dots in Figure \ref{fig:HCO+_clumps} are those which are neither exactly behind nor in front of the clumps. We plotted their values with red points in Figure \ref{fig:twotempB}.

\section{Discussion} \label{sec:diss}

\subsection{Grain alignment with \textit{Purcell torques}}\label{sec:comp}
\citet{1979ApJ...231..404P} proposed three surface processes to drive grains to superathermal rotation. Among these three processes, the formation of $\rm H_{2}$ molecule at the catalytic site was suggested as the dominant mechanism to produce rotational torques. In addition to the work of \citet{1979ApJ...231..404P}, \citet{ 1997ApJ...487..248L}, and \citet{1997ApJ...484..230L} presented the detailed calculations for these torques for a brick-like and a spheroidal grain.  As noted by \citet{2013ApJ...775...84A}; if the $\rm H_{2}$ emission from a region is fluorescent in origin, then the intensity of the emission traces the molecular hydrogen formation rate.   This is because $\rm H_{2}$ photo destruction proceeds via a two step process initiated by a line absorption in the Lyman or Werner bands, followed by a relaxation of the electronic excitation. The direct photodissociation of an $\rm H_{2}$ molecule requires a photon with energy E$>$14.7 eV \citep{1967ApJ...149L..29S}, well above the ionization energy of atomic hydrogen (E=13.6\,eV) and is thus not important in the atomic ISM.  During the relaxation process, the molecule dissociates if the vibrational state of the de-excited molecule has v$>$14. If it re-enters the electronic ground state at a lower vibrational quantum number, the molecule will proceed to the ground state through a rotational-vibrational cascade observable chiefly in the near infrared. Since the branching ratio between dissociation and fluorescence cascade is determined by quantum mechanics, the fluorescence rate is directly proportional to the molecular destruction rate. As the dynamical time scales in a well-established PDR are much longer than those of molecular formation and destruction \citep{tielens2005,2008MNRAS.390.1549M}, the abundance of $\rm H_{2}$ is governed by a detailed balance equation where formation and destruction rates equal each other.  The fluorescent intensities, therefore, trace the formation rate of the molecule.

The formation of molecular hydrogen on the surface of the dust grains helps in spinning up the grains \citep{1979ApJ...231..404P}. When the molecule leaves the grain surface, it imparts a net torque to the dust grain which results in spinning of the grain. This phenomena is called the \enquote{Purcell rocket effect}. These torques cause the grains to rotate supra-thermally (i.e. with rotational energies well above the thermal energies of the gas or dust). These spinning grains interact with interstellar magnetic fields and gets aligned with the field with spinning axes parallel to the field orientation. These \enquote{Purcell torques} play an important role also in modern RAT based grain alignment, where they enhance the radiatively driven alignment allowing a larger fraction of the grains to attain supra-thermal rotation.

%B-G's 2013 work
\citet{2013ApJ...775...84A}, using their spectropolarimetric observation with NOT/ALFOSC and five-band photo-polarimetry with NOT/TURPOL for stars towards IC\,63 studied the correlation between polarization and diffuse $\rm H_{2}$ fluorescence emission in the cloud. They note that the polarization seen in IC\,63 is proportional to the level of $\rm H_{2}$ fluorescence (i.e. $\rm H_{2}$ formation rate) for high intensity $\rm H_{2}$ emission fluorescence (i.e $\rm I_{H_{2}} >10^{-5}\,erg\,cm^{-2}\,s^{-1}\,sr^{-1}$).  They identified this correlation as the signature of enhance dust grain alignment driven by $\rm H_{2}$ formation. No clear correlation was seen for $\rm I_{H_{2}} < 10^{-5}\,erg\,cm^{-2}\,s^{-1}\,sr^{-1}$ suggesting that up to this formation rate, only direct RAT alignment dominates.  They also note a few outliers.  We have used their polarization measurements in addition to values from \citet{2017MNRAS.465..559S} to re-investigate the relation between amount of polarization increment with $\rm H_{2}$ formation on dust grains (see Fig. \ref{fig:PvsH2}). Star numbered 46 (see Fig. \ref{fig:HCO+_clumps}) is in common in the two studies. For this particular star, we use the polarization values from \citet{2017MNRAS.465..559S}\footnote{The sky subtraction process of \citet{2017MNRAS.465..559S} was better for star \#46 as they used CCD-based imaging polarimetry, while TURPOL is a two-channel photo-polarimeter with a fixed sky aperture. \citet{2017MNRAS.465..559S} used the FITSKYPARS task in the IRAF apphot package to subtract the sky background. An aperture is created centered on the star with a radius of ~2 times the FWHM of the Gaussian distribution of the star counts. A sky annulus is created around the star and background contribution is measured within this annulus. The size of the annulus was chosen to yield approximately constant background counts (i.e. avoiding the extended stellar PSF). This value is subtracted from the signal within the aperture. Because of the offset sky-aperture, there are chances of diffused emission contamination in the result of \citet{2013ApJ...775...84A}. They discuss the background contamination in the TURPOL observations in Appendix A.2.}. The polarization results from \citet{2017MNRAS.465..559S} and \citet{2013ApJ...775...84A} are consistent within the uncertainties.

Panel (a) of figure \ref{fig:PvsH2} shows robust linear model (RLM) fitting in distribution of amount of polarization we have from observations and the intensity of $H_{2}$ fluorescence measured at the same locations where we detected polarization. Panel (b) also shows the distribution of same quantities but with linear regression model (LM) fit to the data and uncertainty in the fit model is shown with shaded blue region. These two different models have different function when fit to the data. LM is just ordinary least square fitting which reduces the residuals whereas in RLM, all data points are not treated equally. Robust fitting is not sensitive to outliers but LM is very sensitive to the outlying data points. If the uncertainties are of a fully random nature LM is the more appropriate technique.  However, RLM fitting is useful for samples where additional un-accounted for parameters may be producing significant systematic errors in the data.  The distribution shown in figure \ref{fig:PvsH2} is quite scattered. Therefore, we used both RLM and LM methods. In both the panels, a trend of increase in polarization with intensity of $H_{2}$ fluorescence can be noticed. The correlation is not as good as seen by \citet{2013ApJ...775...84A} who had a lesser number of targets. Although more data should nominally improve the sampling of an underlying relationship, this is only true if the noise in the data set is statistical/random in nature.  If the noise is instead systematic more sampling may not yield tighter correlations. Because of the good correlation seen in \citet{2013ApJ...775...84A}, the formal correlation seen in this larger sample and the highly successful \textit{ab initio} modeling of the alignment in this [physically simple] system by \citet{2015MNRAS.448.1178H}, we feel justified in probing for a \enquote{third parameter} in these data (where RAT alignment and H$_2$ enhanced alignment are the first two).

The most prominent outliers in the plots (here and in \citet{2013ApJ...775...84A}) seem to be associated with large column density regions in the mm-wave maps, especially the compression-ridge region at the side of the nebula facing $\gamma$ Cas.  If the column density can be used as a proxy for space density (i.e. if the cloud can to first order be approximated with a slab geometry) this is where the collision rate would be especially large.  We therefore probe the hypothesis that the outliers represent enhanced collisional disalignment in the material.

\begin{figure}
\centering
\resizebox{13.5cm}{10.0cm}{\includegraphics{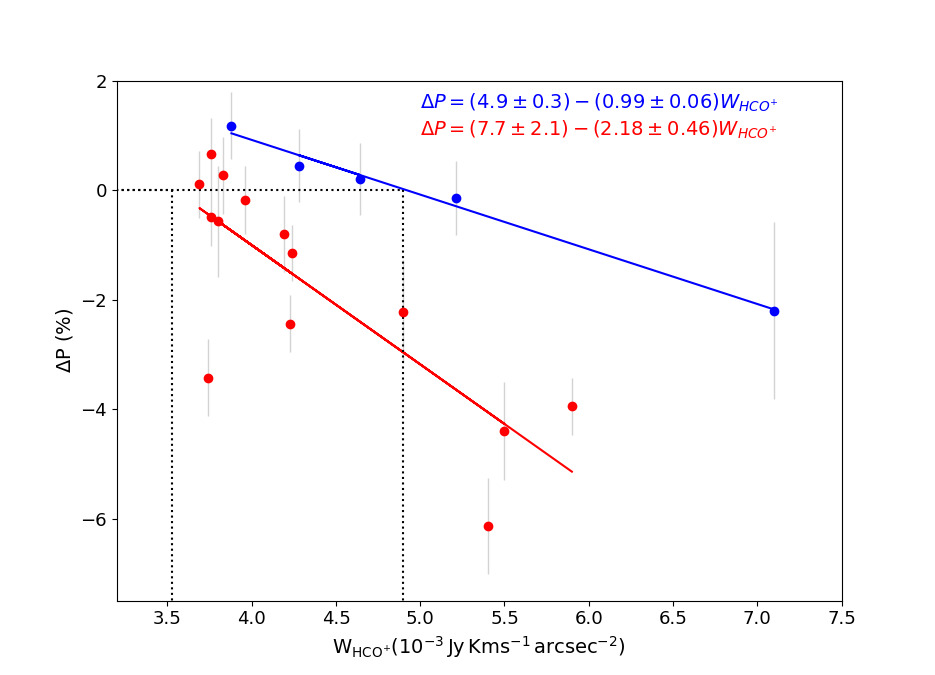}}
\caption{Variation of change in polarization fraction with $\rm HCO^{+}$ integrated flux in the line of sights corresponding to the stars shown in Figure \ref{fig:HCO+_clumps}. The color scheme used for targets in both the plots are consistent. The black dotted lines are $\Delta P$=0 lines corresponding to critical densities of $\rm W_{HCO^{+}}$ (see text for explanation).}\label{fig:twotempB}
\end{figure}

\subsection{Grain-gas collisions and two temperatures in IC\,63}\label{sec:comp}
To probe for a dependence of the measured polarization on the collision rate, we assume that the direct radiative alignment rate of the grains in IC\,63 is close to constant.  This is supported by the fact that the opacity in the gas to photons capable of direct RAT alignment is relatively low as these are only constrained to have $\lambda<2a$ (the opacity to H$_2$ dissociating photons is much smaller than the cloud size since they are much harder with $\lambda<1108$\AA). We then find a nominal correlation with the H$_2$ fluorescent flux to account for the H$_2$ formation enhancement \citep[as done in][]{2013ApJ...775...84A, 2015MNRAS.448.1178H}.  Based on this nominal P vs I(H$_2$) relation we can investigate the possible role of collisions by comparing the deviations from this relationship ($\Delta P$) with the integrated intensity of the $\rm HCO^{+}$ line ($\rm W_{HCO^{+}}$). We use the $\rm HCO^{+}$ column density rather than visual extinction as it has higher accuracy and is assured to probe the IC\,63 cloud only, rather than possible background gas. Other gas tracers, such as CO (J=1-0) or its isotopomers might have been preferable, but are not available at the spatial resolution required to match the pencil beams of the polarimetry.  

The result is shown in Figure \ref{fig:twotempB}. The overall distribution of points shows a weak anti-correlation (with a $\sim$45\% probability based on a \textit{Spearman rank-order} test).  However, a bifurcation in the points is also evident.  To explore this dichotomy, we plot the locations of the polarization lines of sight on a heavily stretched map of $\rm HCO^{+}$ (Fig. \ref{fig:HCO+_clumps}). This map allows us to separate the lines of sight into those that pass through IC\,63 in front of, or behind, a dense clump in the cloud. We have marked the former with red symbols in both Figures \ref{fig:HCO+_clumps} and \ref{fig:twotempB} and the later with blue symbols. A number of lines of sight do not clearly fall in either category which are plotted as white symbols in Figure \ref{fig:HCO+_clumps}. If the distinguishing separations between the different lines of sight is the illumination of the gas (and dust) by $\gamma$ Cas, then these latter points shown with white color should be more closely related to the ones in front of a clump (red symbols), as the distinguishing characteristic is whether the radiation from the background star has been damped or not by the dense gas.

Considering the color scheme in Figure \ref{fig:HCO+_clumps}, we notice a clear separation of the full sample into two populations. The populations with blue symbols which are in shaded region behind $\rm HCO^{+}$ clumps (Figure \ref{fig:HCO+_clumps}), show significantly shallower slope (--0.99$\pm$0.06) than that of the population with the red symbols (--2.18$\pm$0.46) which are in front of the $\rm HCO^{+}$ clumps (see Figure \ref{fig:twotempB}). The two slopes here suggest different rates of drop in amount of polarization with the gas density.

The anti-correlation seen in the distribution is expected for the collisional disalignment, whose rate (C) is proportional to the gas density ($\rm n_{gas}$) and gas velocity ($\rm v_{gas}$) 
\begin{equation}
    C\propto n_{gas} \cdot \sqrt{T_{gas}}
\end{equation}
\noindent
where, for thermal motion, the gas temperature (T$_{gas}$) is related to average velocity of the gas (v$_{gas}$) by relation {\bf $T_{gas} = \frac{\mu m_{H}}{k_{B}}v^2_{gas}$}.  As discussed above, we are here applying the local density variations in a perturbation mode around the steady-state solution described by equation (\ref{balanceequ}), which means that $\Delta P$ vs. n$_{gas}$ should show linear anti correlation with a slope proportional to the square root of the gas temperature.  If we, as suggested above, approximate the cloud as a slab geometry, then the integrated $\rm HCO^{+}$ intensity can be used as a proxy for gas density. Figure \ref{fig:twotempB} shows two separate slopes fit to the two populations of targets shown in red and blue symbols with the point tracing more intensely illuminated gas following a steeper plot in $\rm \Delta{P}$ vs. $\rm W_{HCO^{+}}$. This would be expected if the gas is heated by the radiation from the star and gas \textit{in the shade} is cooler than the gas in the direct illumination of $\rm \gamma\ $Cas radiation. Similar to the opacity differences between H$_2$ photodissociation and the subsequent re-formation, the photons responsible for heating the gas, through photoelectric emission from small grains will likely also have a higher effective opacity into the cloud than those responsible for the direct RAT alignment. This assumes that the small grain population is present in both \enquote{hot} and \enquote{cool} gas, which is likely as the clumps are not dense or dark enough to support significant grain growth \citep{Ysard2013, Vaillancourt2020}. This depends on the poorly known work function of the grain material, but if we assume a work function of 4.4 eV (equivalent to a photon wavelength of $\lambda=2820$\AA) for graphite (PAHs) and 8 eV ($\Leftrightarrow \lambda=1550$\AA) for silicates  \citep{2001ApJS..134..263W}, then also the gas heating is likely to be more spatially structured that the direct RAT alignment, and it is reasonable to thence separate the alignment from the disalignment dependence of the radiation field. 

To probe the [dis-]alignment efficiency, the ideal probe would be the normalized fractional polarization ($P/A_{V}$), which removes changes due to total column density.  However, while we have some $A_{V}$ measurements of the targets shown in Figure \ref{fig:twotempB} some line of sight extinction measurements are fully missing, while for other the existing ones are of marginal accuracy.  Therefore, performing the analysis in $P/A_{V}$ introduces additional uncertainties. Doing this analysis shows that the parameters of the two slopes in this figure are within the mutual uncertainties of those utilizing the normalized fractional polarization $P/A_{V}$, and implies that the approximation of the cloud as a plane-parallel slab is reasonable.

The two slopes in Figure \ref{fig:twotempB} are consistent with the two temperature hypothesis of \citet{2009MNRAS.400..622T}. The ratio of the square roots of the two temperatures 106\,K and 685\,K of warm and hot regions, respectively, in Figure \ref{Fig:thi09} is 2.54$\pm$1.83. The ratio of the two slopes obtained in Figure \ref{fig:twotempB} is 2.20$\pm$0.49. These values are consistent within their joint uncertainty of $\sim$1.89.  The higher rate of gas-grain collision in the hot regions leads to higher disalignment rate of the aligned dust grains and hence stronger decrements in polarization, as seen in Figure \ref{fig:twotempB}.

The steady-state solution of the alignment (equation (\ref{balanceequ})) has a 1/(n$_{gas}\cdot \sqrt{T_{gas}}$) dependence in the RAT term, and, therefore, the $\Delta$P=0 location should depend on these parameters. Figure \ref{fig:twotempB} shows dotted lines plotted for $\Delta P$=0 and the corresponding two integrated intensities of $\rm HCO^{+}$ on x-axis. This \enquote{critical density} for the hotter gas (toward higher temperature) occurs at a lower value than for the colder gas as would be expected from the above dependence. This supports our conclusion that the two sub-sets and their associated linear anti-correlations represent the two - warm and hot - phases of the gas in IC\,63 PDR. It is possible to an unknown degree that the rotational disruption by RAT (RATD; \citealt{hoang2019}) may play a minor role in affecting the critical density.

It is worth noting that a dual temperature distribution in the molecular hydrogen line excitation is a well known phenomenon in UV observations of diffuse gas.  There, the J=1, J=0 excitation shows a temperature of typically $\sim$80--130\,K \citep{2015MNRAS.446.2490N} consistent with the cool neutral medium. For J=3 and higher, a typically much higher excitation temperature is derived $\sim$500-1000\,K \citep{Shaw2005}. As shown by \citet{1990ApJ...359L..19L}, the column density in the high J-levels is correlated with that of CH$^+$.  Since the formation path of this radical requires an endothermic reaction with an activation energy exceeding 4600\,K:
\begin{equation}
    C^+ + H_2 \rightarrow CH^+ + H - 4640\,K
\end{equation}
 \citep[][and refs. therein]{1986MNRAS.220..801P}, this implies that some gas along the line of sight either is now in warm/hot regions or recently was.  Hence the two temperature excitation seen in UV observations likely corresponds to distinct thermal components along the line of sight. Our proposed separation of thermal components corresponds to a similar structure, albeit in the plane of the sky.

\section{Summary and conclusions}
We have combined optical polarization observations with $\rm H_{2}$ and $\rm HCO^{+}$ line observations towards the IC\,63 cloud in order to investigate the effects on grain alignment efficiency and polarization values due to gas-grain collisions. We find that, assuming a close to constant direct radiative alignment rate (as suggested by \citet{2015MNRAS.448.1178H}) enhanced by H$_2$ formation torques, the polarization is decreased by gas-grain collisions consistent with a two temperature structure, where the higher temperature corresponds to line of sight probing gas illuminated by the un-attenuated radiation from $\gamma$\ Cas, and the lower temperature corresponds to gas shaded by dense clumps in the cloud. The ratio of the two slopes is consistent with the two-temperature solution seen in pure rotational excitation observations of H$_2$ \citep{2009MNRAS.400..622T}.  We find that relative critical gas density at which the gas-grain collisions are sufficiently effective to disalign the grains is smaller in the hot gas than in the warm gas, supporting our conclusion.  Further observations, including direct measurements of the gas space density, are required to confirm these conclusions.

\bigskip
\bigskip
\bigskip
We thank anonymous referee for helping us to considerably improve the content of the paper. We gratefully acknowledge the generous telescope allocations by the review committees of the NOT and CFHT and the expert support by the observatory staffs. The dataset presented here were obtained in part using the Nordic Optical Telescope (NOT) with ALFOSC, which is provided by the Instituto de Astrofisica de Andalucia (IAA) under a joint agreement with the University of Copenhagen and NOTSA. Based also, in part, on observations obtained with WIRCam, a joint project of the Canada-France-Hawaii Telescope (CFHT), Taiwan, Korea, Canada, France. The CFHT is operated by the National Research Council (NRC) of Canada, the Institute National des Sciences de l’Univers of the Centre National de la Recherche Scientifique of France, and the University of Hawaii. B-G.A. and A.S. acknowledge financial support from the NSF through grant AST-1715876. J.E.V. acknowledges financial support from NASA through award number SOF 05-0038 issued by USRA. We thank Adam Mantz and Maria Jose Maureira for CARMA observations.

\facility{CFHT, CARMA, WHT, AIMPOL, IRAM}


\software{Astropy \citep{2013A&A...558A..33A}, Matplotlib \citep{Hunter2007}, SciPy \citep{2020SciPy-NMeth}, NumPy \citep{2020NumPy-Array}}

%\end{document}

\bibliography{ref}{}

\begin{thebibliography}{}
\expandafter\ifx\csname natexlab\endcsname\relax\def\natexlab#1{#1}\fi
\providecommand{\url}[1]{\href{#1}{#1}}
\providecommand{\dodoi}[1]{doi:~\href{http://doi.org/#1}{\nolinkurl{#1}}}
\providecommand{\doeprint}[1]{\href{http://ascl.net/#1}{\nolinkurl{http://ascl.net/#1}}}
\providecommand{\doarXiv}[1]{\href{https://arxiv.org/abs/#1}{\nolinkurl{https://arxiv.org/abs/#1}}}

\bibitem[{{Andersson} {et~al.}(2015){Andersson}, {Lazarian}, \&
  {Vaillancourt}}]{2015ARA&A..53..501A}
{Andersson}, B.~G., {Lazarian}, A., \& {Vaillancourt}, J.~E. 2015, \araa, 53,
  501, \dodoi{10.1146/annurev-astro-082214-122414}

\bibitem[{{Andersson} {et~al.}(2011){Andersson}, {Pintado}, {Potter},
  {Strai{\v{z}}ys}, \& {Charcos-Llorens}}]{2011A&A...534A..19A}
{Andersson}, B.~G., {Pintado}, O., {Potter}, S.~B., {Strai{\v{z}}ys}, V., \&
  {Charcos-Llorens}, M. 2011, \aap, 534, A19,
  \dodoi{10.1051/0004-6361/201117670}

\bibitem[{{Andersson} \& {Potter}(2007)}]{2007ApJ...665..369A}
{Andersson}, B.~G., \& {Potter}, S.~B. 2007, \apj, 665, 369,
  \dodoi{10.1086/519755}

\bibitem[{{Andersson} \& {Potter}(2010)}]{2010ApJ...720.1045A}
---. 2010, \apj, 720, 1045, \dodoi{10.1088/0004-637X/720/2/1045}

\bibitem[{{Andersson} {et~al.}(2013){Andersson}, {Piirola}, {De Buizer},
  {Clemens}, {Uomoto}, {Charcos-Llorens}, {Geballe}, {Lazarian}, {Hoang}, \&
  {Vornanen}}]{2013ApJ...775...84A}
{Andersson}, B.~G., {Piirola}, V., {De Buizer}, J., {et~al.} 2013, \apj, 775,
  84, \dodoi{10.1088/0004-637X/775/2/84}

\bibitem[{{Andrews} {et~al.}(2018){Andrews}, {Peeters}, {Tielens}, \&
  {Okada}}]{2018A&A...619A.170A}
{Andrews}, H., {Peeters}, E., {Tielens}, A.~G.~G.~M., \& {Okada}, Y. 2018,
  \aap, 619, A170, \dodoi{10.1051/0004-6361/201832808}

\bibitem[{{Astropy Collaboration} {et~al.}(2013){Astropy Collaboration},
  {Robitaille}, {Tollerud}, {Greenfield}, {Droettboom}, {Bray}, {Aldcroft},
  {Davis}, {Ginsburg}, {Price-Whelan}, {Kerzendorf}, {Conley}, {Crighton},
  {Barbary}, {Muna}, {Ferguson}, {Grollier}, {Parikh}, {Nair}, {Unther},
  {Deil}, {Woillez}, {Conseil}, {Kramer}, {Turner}, {Singer}, {Fox}, {Weaver},
  {Zabalza}, {Edwards}, {Azalee Bostroem}, {Burke}, {Casey}, {Crawford},
  {Dencheva}, {Ely}, {Jenness}, {Labrie}, {Lim}, {Pierfederici}, {Pontzen},
  {Ptak}, {Refsdal}, {Servillat}, \& {Streicher}}]{2013A&A...558A..33A}
{Astropy Collaboration}, {Robitaille}, T.~P., {Tollerud}, E.~J., {et~al.} 2013,
  \aap, 558, A33, \dodoi{10.1051/0004-6361/201322068}

\bibitem[{Bradley {et~al.}(2019)Bradley, Sip{\H o}cz, Robitaille, Tollerud,
  Vin{\'{\i}}cius, Deil, Barbary, G{\"u}nther, Cara, Busko, Conseil,
  Droettboom, Bostroem, Bray, Bratholm, Wilson, Craig, Barentsen, Pascual,
  Donath, Greco, Perren, Lim, \& Kerzendorf}]{Bradley_2019_2533376}
Bradley, L., Sip{\H o}cz, B., Robitaille, T., {et~al.} 2019, astropy/photutils:
  v0.6, \dodoi{10.5281/zenodo.2533376}

\bibitem[{{Chandrasekhar} \& {Fermi}(1953)}]{1953ApJ...118..113C}
{Chandrasekhar}, S., \& {Fermi}, E. 1953, \apj, 118, 113,
  \dodoi{10.1086/145731}

\bibitem[{{Crutcher}(2004)}]{2004Ap&SS.292..225C}
{Crutcher}, R.~M. 2004, \apss, 292, 225,
  \dodoi{10.1023/B:ASTR.0000045021.42255.95}

\bibitem[{{Davis} \& {Greenstein}(1951)}]{1951ApJ...114..206D}
{Davis}, Leverett, J., \& {Greenstein}, J.~L. 1951, \apj, 114, 206,
  \dodoi{10.1086/145464}

\bibitem[{{Davis}(1951)}]{Davis1951}
{Davis}, L. 1951, Physical Review, 81, 890, \dodoi{10.1103/PhysRev.81.890.2}

\bibitem[{{Dolginov} \& {Mitrofanov}(1976)}]{1976Ap&SS..43..291D}
{Dolginov}, A.~Z., \& {Mitrofanov}, I.~G. 1976, \apss, 43, 291,
  \dodoi{10.1007/BF00640010}

\bibitem[{{Draine} \& {Lazarian}(1998)}]{1998ApJ...508..157D}
{Draine}, B.~T., \& {Lazarian}, A. 1998, \apj, 508, 157, \dodoi{10.1086/306387}

\bibitem[{{Draine} \& {Weingartner}(1996)}]{1996ApJ...470..551D}
{Draine}, B.~T., \& {Weingartner}, J.~C. 1996, \apj, 470, 551,
  \dodoi{10.1086/177887}

\bibitem[{{Federman} {et~al.}(1979){Federman}, {Glassgold}, \&
  {Kwan}}]{federman1979}
{Federman}, S.~R., {Glassgold}, A.~E., \& {Kwan}, J. 1979, \apj, 227, 466,
  \dodoi{10.1086/156753}

\bibitem[{{Fleming} {et~al.}(2010){Fleming}, {France}, {Lupu}, \&
  {McCandliss}}]{2010ApJ...725..159F}
{Fleming}, B., {France}, K., {Lupu}, R.~E., \& {McCandliss}, S.~R. 2010, \apj,
  725, 159, \dodoi{10.1088/0004-637X/725/1/159}

\bibitem[{{France} {et~al.}(2005){France}, {Andersson}, {McCandliss}, \&
  {Feldman}}]{2005ApJ...628..750F}
{France}, K., {Andersson}, B.~G., {McCandliss}, S.~R., \& {Feldman}, P.~D.
  2005, \apj, 628, 750, \dodoi{10.1086/430878}

\bibitem[{Harris {et~al.}(2020)Harris, Millman, van~der Walt, Gommers,
  Virtanen, Cournapeau, Wieser, Taylor, Berg, Smith, Kern, Picus, Hoyer, van
  Kerkwijk, Brett, Haldane, Fernández~del Río, Wiebe, Peterson,
  Gérard-Marchant, Sheppard, Reddy, Weckesser, Abbasi, Gohlke, \&
  Oliphant}]{2020NumPy-Array}
Harris, C.~R., Millman, K.~J., van~der Walt, S.~J., {et~al.} 2020, Nature, 585,
  357–362, \dodoi{10.1038/s41586-020-2649-2}

\bibitem[{{Hoang} {et~al.}(2015){Hoang}, {Lazarian}, \&
  {Andersson}}]{2015MNRAS.448.1178H}
{Hoang}, T., {Lazarian}, A., \& {Andersson}, B.~G. 2015, \mnras, 448, 1178,
  \dodoi{10.1093/mnras/stu2758}

\bibitem[{{Hoang} {et~al.}(2019){Hoang}, {Tram}, {Lee}, \& {Ahn}}]{hoang2019}
{Hoang}, T., {Tram}, L.~N., {Lee}, H., \& {Ahn}, S.-H. 2019, Nature Astronomy,
  3, 766, \dodoi{10.1038/s41550-019-0763-6}

\bibitem[{{Hough} {et~al.}(2008){Hough}, {Aitken}, {Whittet}, {Adamson}, \&
  {Chrysostomou}}]{hough2008}
{Hough}, J.~H., {Aitken}, D.~K., {Whittet}, D.~C.~B., {Adamson}, A.~J., \&
  {Chrysostomou}, A. 2008, \mnras, 387, 797,
  \dodoi{10.1111/j.1365-2966.2008.13274.x}

\bibitem[{Hunter(2007)}]{Hunter2007}
Hunter, J.~D. 2007, Computing in Science \& Engineering, 9, 90,
  \dodoi{10.1109/MCSE.2007.55}

\bibitem[{{Jansen} {et~al.}(1994){Jansen}, {van Dishoeck}, \&
  {Black}}]{1994A&A...282..605J}
{Jansen}, D.~J., {van Dishoeck}, E.~F., \& {Black}, J.~H. 1994, \aap, 282, 605

\bibitem[{{Jansen} {et~al.}(1995){Jansen}, {van Dishoeck}, {Black}, {Spaans},
  \& {Sosin}}]{1995A&A...302..223J}
{Jansen}, D.~J., {van Dishoeck}, E.~F., {Black}, J.~H., {Spaans}, M., \&
  {Sosin}, C. 1995, \aap, 302, 223

\bibitem[{{Jones} \& {Spitzer}(1967)}]{jones1967}
{Jones}, R.~V., \& {Spitzer}, Lyman, J. 1967, \apj, 147, 943,
  \dodoi{10.1086/149086}

\bibitem[{{Jones} {et~al.}(1984){Jones}, {Hyland}, \& {Bailey}}]{jones1984}
{Jones}, T.~J., {Hyland}, A.~R., \& {Bailey}, J. 1984, \apj, 282, 675,
  \dodoi{10.1086/162247}

\bibitem[{{Karr} {et~al.}(2005){Karr}, {Noriega-Crespo}, \&
  {Martin}}]{2005AJ....129..954K}
{Karr}, J.~L., {Noriega-Crespo}, A., \& {Martin}, P.~G. 2005, \aj, 129, 954,
  \dodoi{10.1086/426912}

\bibitem[{{Kaufman} {et~al.}(1999){Kaufman}, {Wolfire}, {Hollenbach}, \&
  {Luhman}}]{1999ApJ...527..795K}
{Kaufman}, M.~J., {Wolfire}, M.~G., {Hollenbach}, D.~J., \& {Luhman}, M.~L.
  1999, \apj, 527, 795, \dodoi{10.1086/308102}

\bibitem[{{Lambert} {et~al.}(1990){Lambert}, {Sheffer}, \&
  {Crane}}]{1990ApJ...359L..19L}
{Lambert}, D.~L., {Sheffer}, Y., \& {Crane}, P. 1990, \apjl, 359, L19,
  \dodoi{10.1086/185786}

\bibitem[{{Lazarian} \& {Draine}(1997)}]{1997ApJ...487..248L}
{Lazarian}, A., \& {Draine}, B.~T. 1997, \apj, 487, 248, \dodoi{10.1086/304587}

\bibitem[{{Lazarian} \& {Draine}(1999)}]{1999ApJ...520L..67L}
---. 1999, \apjl, 520, L67, \dodoi{10.1086/312137}

\bibitem[{{Lazarian} \& {Hoang}(2007)}]{2007MNRAS.378..910L}
{Lazarian}, A., \& {Hoang}, T. 2007, \mnras, 378, 910,
  \dodoi{10.1111/j.1365-2966.2007.11817.x}

\bibitem[{{Lazarian} \& {Roberge}(1997)}]{1997ApJ...484..230L}
{Lazarian}, A., \& {Roberge}, W.~G. 1997, \apj, 484, 230,
  \dodoi{10.1086/304309}

\bibitem[{{Mathis}(1986)}]{1986ApJ...308..281M}
{Mathis}, J.~S. 1986, \apj, 308, 281, \dodoi{10.1086/164499}

\bibitem[{{Medan} \& {Andersson}(2019)}]{medan2019}
{Medan}, I., \& {Andersson}, B.-G. 2019, \apj, 873, 87,
  \dodoi{10.3847/1538-4357/ab063c}

\bibitem[{{Morata} \& {Herbst}(2008)}]{2008MNRAS.390.1549M}
{Morata}, O., \& {Herbst}, E. 2008, \mnras, 390, 1549,
  \dodoi{10.1111/j.1365-2966.2008.13837.x}

\bibitem[{{Naslim} {et~al.}(2015){Naslim}, {Kemper}, {Madden}, {Hony}, {Chu},
  {Galliano}, {Bot}, {Yang}, {Seok}, {Oliveira}, {van Loon}, {Meixner}, {Li},
  {Hughes}, {Gordon}, {Otsuka}, {Hirashita}, {Morata}, {Lebouteiller},
  {Indebetouw}, {Srinivasan}, {Bernard}, \& {Reach}}]{2015MNRAS.446.2490N}
{Naslim}, N., {Kemper}, F., {Madden}, S.~C., {et~al.} 2015, \mnras, 446, 2490,
  \dodoi{10.1093/mnras/stu2276}

\bibitem[{{Pineau des Forets} {et~al.}(1986){Pineau des Forets}, {Flower},
  {Hartquist}, \& {Dalgarno}}]{1986MNRAS.220..801P}
{Pineau des Forets}, G., {Flower}, D.~R., {Hartquist}, T.~W., \& {Dalgarno}, A.
  1986, \mnras, 220, 801, \dodoi{10.1093/mnras/220.4.801}

\bibitem[{{Polehampton} {et~al.}(2005){Polehampton}, {Wyrowski}, \&
  {Schilke}}]{2005IAUS..231P.148P}
{Polehampton}, E.~T., {Wyrowski}, F., \& {Schilke}, P. 2005, in IAU Symposium,
  Vol. 231, Astrochemistry: Recent Successes and Current Challenges, ed. D.~C.
  {Lis}, G.~A. {Blake}, \& E.~{Herbst}, 148

\bibitem[{{Puget} {et~al.}(2004){Puget}, {Stadler}, {Doyon}, {Gigan},
  {Thibault}, {Luppino}, {Barrick}, {Benedict}, {Forveille}, {Rambold},
  {Thomas}, {Vermeulen}, {Ward}, {Beuzit}, {Feautrier}, {Magnard}, {Mella},
  {Preis}, {Vallee}, {Wang}, {Lin}, {Hall}, \& {Hodapp}}]{2004SPIE.5492..978P}
{Puget}, P., {Stadler}, E., {Doyon}, R., {et~al.} 2004, Society of
  Photo-Optical Instrumentation Engineers (SPIE) Conference Series, Vol. 5492,
  {WIRCam: the infrared wide-field camera for the Canada-France-Hawaii
  Telescope}, ed. A.~F.~M. {Moorwood} \& M.~{Iye}, 978--987,
  \dodoi{10.1117/12.551097}

\bibitem[{{Purcell}(1979)}]{1979ApJ...231..404P}
{Purcell}, E.~M. 1979, \apj, 231, 404, \dodoi{10.1086/157204}

\bibitem[{{R{\"o}llig} {et~al.}(2007){R{\"o}llig}, {Abel}, {Bell}, {Bensch},
  {Black}, {Ferland}, {Jonkheid}, {Kamp}, {Kaufman}, {Le Bourlot}, {Le Petit},
  {Meijerink}, {Morata}, {Ossenkopf}, {Roueff}, {Shaw}, {Spaans}, {Sternberg},
  {Stutzki}, {Thi}, {van Dishoeck}, {van Hoof}, {Viti}, \&
  {Wolfire}}]{2007A&A...467..187R}
{R{\"o}llig}, M., {Abel}, N.~P., {Bell}, T., {et~al.} 2007, \aap, 467, 187,
  \dodoi{10.1051/0004-6361:20065918}

\bibitem[{{Sault} \& {Noordam}(1995)}]{1995Sault}
{Sault}, R.~J., \& {Noordam}, J.~E. 1995, \aaps, 109, 593

\bibitem[{{Shaw} {et~al.}(2005){Shaw}, {Ferland}, {Abel}, {Stancil}, \& {van
  Hoof}}]{Shaw2005}
{Shaw}, G., {Ferland}, G.~J., {Abel}, N.~P., {Stancil}, P.~C., \& {van Hoof},
  P.~A.~M. 2005, \apj, 624, 794, \dodoi{10.1086/429215}

\bibitem[{{Soam} {et~al.}(2017){Soam}, {Maheswar}, {Lee}, {Neha}, \&
  {Andersson}}]{2017MNRAS.465..559S}
{Soam}, A., {Maheswar}, G., {Lee}, C.~W., {Neha}, S., \& {Andersson}, B.~G.
  2017, \mnras, 465, 559, \dodoi{10.1093/mnras/stw2649}

\bibitem[{{Stecher} \& {Williams}(1967)}]{1967ApJ...149L..29S}
{Stecher}, T.~P., \& {Williams}, D.~A. 1967, \apjl, 149, L29,
  \dodoi{10.1086/180047}

\bibitem[{{Thi} {et~al.}(2009){Thi}, {van Dishoeck}, {Bell}, {Viti}, \&
  {Black}}]{2009MNRAS.400..622T}
{Thi}, W.~F., {van Dishoeck}, E.~F., {Bell}, T., {Viti}, S., \& {Black}, J.
  2009, \mnras, 400, 622, \dodoi{10.1111/j.1365-2966.2009.15501.x}

\bibitem[{{Thi} {et~al.}(1997){Thi}, {van Dishoeck}, {Jansen}, {Spaans}, {Li},
  {Evans}, \& {Jaffe}}]{1997ESASP.419..299T}
{Thi}, W.~F., {van Dishoeck}, E.~F., {Jansen}, D.~J., {et~al.} 1997, in ESA
  Special Publication, Vol. 419, The first ISO workshop on Analytical
  Spectroscopy, ed. A.~M. {Heras}, K.~{Leech}, N.~R. {Trams}, \& M.~{Perry},
  299

\bibitem[{{Tielens}(2005)}]{tielens2005}
{Tielens}, A.~G.~G.~M. 2005, {The Physics and Chemistry of the Interstellar
  Medium} (Cambridge, UK: Cambridge University Press)

\bibitem[{{Troland} \& {Crutcher}(2008)}]{2008ApJ...680..457T}
{Troland}, T.~H., \& {Crutcher}, R.~M. 2008, \apj, 680, 457,
  \dodoi{10.1086/587546}

\bibitem[{{Vaillancourt} {et~al.}(2020){Vaillancourt}, {Andersson}, {Clemens},
  {Piirola}, {Hoang}, {Becklin}, \& {Caputo}}]{Vaillancourt2020}
{Vaillancourt}, J.~E., {Andersson}, B.-G., {Clemens}, D.~P., {et~al.} 2020,
  arXiv e-prints, arXiv:2011.00114.
\newblock \doarXiv{2011.00114}

\bibitem[{{van Dishoeck} \& {Black}(1986)}]{1986ApJS...62..109V}
{van Dishoeck}, E.~F., \& {Black}, J.~H. 1986, \apjs, 62, 109,
  \dodoi{10.1086/191135}

\bibitem[{{van Leeuwen}(2007)}]{vanleeuwen2007}
{van Leeuwen}, F. 2007, \aap, 474, 653, \dodoi{10.1051/0004-6361:20078357}

\bibitem[{Virtanen {et~al.}(2020)Virtanen, Gommers, Oliphant, Haberland, Reddy,
  Cournapeau, Burovski, Peterson, Weckesser, Bright, {van der Walt}, Brett,
  Wilson, Millman, Mayorov, Nelson, Jones, Kern, Larson, Carey, Polat, Feng,
  Moore, {VanderPlas}, Laxalde, Perktold, Cimrman, Henriksen, Quintero, Harris,
  Archibald, Ribeiro, Pedregosa, {van Mulbregt}, \& {SciPy 1.0
  Contributors}}]{2020SciPy-NMeth}
Virtanen, P., Gommers, R., Oliphant, T.~E., {et~al.} 2020, Nature Methods, 17,
  261, \dodoi{10.1038/s41592-019-0686-2}

\bibitem[{{Weingartner} \& {Draine}(2001)}]{2001ApJS..134..263W}
{Weingartner}, J.~C., \& {Draine}, B.~T. 2001, \apjs, 134, 263,
  \dodoi{10.1086/320852}

\bibitem[{{Whittet}(2003)}]{whittet2003}
{Whittet}, D. C.~B. 2003, Dust in the galactic environment - 2:nd ed. (Dust in
  the galactic environment Institute of Physics Publishing, 390 p.).
\newblock
  \url{http://adsabs.harvard.edu/cgi-bin/nph-bib_query?bibcode=2003QB791.W45......&db_key=AST}

\bibitem[{{Whittet} {et~al.}(2008){Whittet}, {Hough}, {Lazarian}, \&
  {Hoang}}]{2008ApJ...674..304W}
{Whittet}, D.~C.~B., {Hough}, J.~H., {Lazarian}, A., \& {Hoang}, T. 2008, \apj,
  674, 304, \dodoi{10.1086/525040}

\bibitem[{{Ysard} {et~al.}(2013){Ysard}, {Abergel}, {Ristorcelli}, {Juvela},
  {Pagani}, {K{\"o}nyves}, {Spencer}, {White}, \& {Zavagno}}]{Ysard2013}
{Ysard}, N., {Abergel}, A., {Ristorcelli}, I., {et~al.} 2013, \aap, 559, A133,
  \dodoi{10.1051/0004-6361/201322066}

\end{thebibliography}
\bibliographystyle{aasjournal}

\end{document}